\documentclass[10pt,a4paper]{article}
\usepackage[utf8]{inputenc}
\usepackage{amsmath}
\usepackage{amsfonts}
\usepackage{amssymb}
\usepackage{graphicx}
\date{}

\newcommand{\am }{{\footnotesize AMPT }}
\newcommand{\amm}{{\footnotesize AMPT}}

\begin{document}
\title{Event-by-Event Fluctuations Clusterization and Entropy Production in AA Collisions at AGS and SPS Energies}
\author{Bushra Ali, Shaista Khan, Anuj Chandra and Shakeel Ahmad\footnote{Shakeel.Ahmad@cern.ch}}

\maketitle

\begin{abstract}
\noindent Event-by-event (ebe) fluctuations in mean pseudorapidity values of relativistic charged particles in full phase space is studied by analysing experimental data on $^{16}O-AgBr$ collisions at 14.5A, 60A, and 200A GeV/c and $^{32}S-AgBr$ collisions at 200A GeV/c. The findings are compared with the prediction of A Multi-Phase Transport(\amm) model and those obtained from the analysis of correlation free Monte-Carlo events. Fluctuations in mean pseudorapidity distributions are noticed to be in excess to that expected from the statistically independent particle emission. The observed dependence of the fluctuation strength measure parameter,$\phi$ on the beam energy and number of participating target nucleons indicate that nucleus-nucleus collisions can not be treated as simple superposition of multiple nucleon-nucleon interactions. Presence of clusters or jet-like phenomena in multihadron final states are searched for on ebe basis by using the concept of Jaynes Shannon entropy. The findings indicate the presence of cluster like objects in the experimental data with their size and frequency increasing with increasing beam energy. These observations, in turn suggest that the clustering or jet-like algorithm adopted in the present study may be used as a tool for triggering different classes of events.
\end{abstract}

\section{Introduction}
\noindent Any physical quantity measured in an experiment is subject to fluctuations. These fluctuations depend on the property of the system and provide useful information about the system under investigation\cite{1,2}. As regards nucleus-nucleus (AA) collisions, the system created is considered as a dense and hot fireball consisting of partonic and(or) hadronic matter\cite{1}. One of the main goals of such a study is to investigate the existence of partonic matter in the early life of a fireball\cite{2}. Investigations involving fluctuations in AA collisions are expected to help check the idea that the fluctuations of a thermal system are clearly related to various susceptibilities\cite{1,2,3} and may serve as an indicator of the possible phase transitions. Moreover, large event-by-event(ebe) fluctuations, if present, might be taken as a signal for its formation\cite{3,4,5}. A key problem in the search of QGP is to identify the QGP signatures by studying the experimental observables\cite{6}. In AA collisions, if the system undergoes a phase transition from hadronic matter to QGP, the degrees of freedom in the two phases would be quite different\cite{6}. Due to the large difference, correlations and fluctuations of thermodynamic quantities and (or) the distributions of the produced charged particles in a phase space may change, apparently lacking any definite pattern. It is, therefore, required that analysis of the data involving such collisions should be carried out on ebe basis. \\

\noindent A major contribution to the observed fluctuations results due to finite event multiplicities. These fluctuations are referred to as the statistical fluctuations and can be estimated by assuming the independent emission of particles or by using event mixing techniques\cite{2,7}. The other fluctuations are envisaged to be of dynamical origin and may be divided into two groups\cite{1}, i) fluctuations which do not change on ebe basis, for instance two particle correlations arising due to the decay of resonances or Bose Einstein statistics and ii) fluctuations which vary on ebe basis, for example charged to neutral particle ratio due to creation of DCC (Disoriented chiral condensate) region, or production of jets which contribute to the high transverse momentum ($p_{T}$) tail of $p_{T}$ distribution. DCC is a region in space in which chiral order parameter points in a direction in isospin space, which is different from that favoured by the true vacuum\cite{2,8}. It has been argued by Bjorken et al.\cite{9,10} that the pion field is oriented along a single direction in isospin space throughout a large fraction of volume of the colliding system, referred to as the DCC region. This results in the production of a spectacular event structure having some regions of the detector dominated by charged pions while the other by neutral pions. This type of behaviour may have been observed in Centauro events\cite{11}. It would be interesting to check whether there exists mechanism by which DCC formation in AA collisions may be explained. If such mechanism invoke the QCD phase transition in an essential way, then signals of DCC formation may help drawing conclusions regarding QGP production. \\

\noindent ebe fluctuations in AA collisions at SPS, RHIC, and LHC energies have been investigated by several workers\cite{12,13,14,15,16,17,18,19}. Various hadronic observables, produced in central $^{208}$Pb -$^{208}$Pb collisions have been found\cite{14,15} to exhibit qualitative changes in their energy dependence, in the SPS energy range. A comparison of these observables with the predictions of statistical or(and) hadronic transport model(s) suggests that the experimental results are consistent with the expected signals of the onset of a phase transition in AA collisions at SPS energies\cite{14,15,20}.\\

\noindent It has been proposed\cite{7,21} that studying the deviation of the distributions of ebe mean transverse momentum ($M_{p_{T}}$) or mean pseudorapidity ($M_{\eta}$) from the random distributions expected from statistically independent emission, would provide interesting information about the randomization or thermalization characteristics of high multiplicity events produced in AA collisions. A measure of fluctuations, $\Phi$ has been defined by Gazdzicki and Mrowczynski\cite{21} which vanishes in case of independent emissions of particles from a single source. However if AA collisions are taken to be as incoherent superposition of nucleon-nucleon (\(nn\)) collisions, the values should be independent of \(nn\) subprocesses and match with that obtained for \(nn\) collisions\cite{22}.\\

\noindent An attempt is, therefore, made to study the ebe fluctuation in terms of $M_{n}$ and $\Phi_{\eta}$ by analysing the experimental data on AA collisions in a wide range of incident energy and with varying system size. Presence of jet-like phenomena and entropy production has also been looked into to check whether the high particle density regions arise due to the decay of a heavier cluster or several clusters or jets of relatively smaller sizes\cite{2,22}.\\

\section{Details of the data}
\noindent Four samples of events produced in the interactions of $^{16}$O and $^{32}$S beams with AgBr group of nuclie in emulsion has been used. Details of the data samples used in the analysis are listed in Table$~$1. These events are taken from the series of experiments carried out by EMU01 collaboration\cite{23,24,25,26}. All the relevant information about the data, like selection of events, track classification, extraction of AgBr group of events, method of measuring the angles of relativistic charged particles, etc., may be found elsewhere\cite{2,6,15,17,27,28}.  \\

\begin{footnotesize}
\begin{table}
\begin{center}
\begin{tabular}{ccc} \hline
Energy & Type of & No. of   \\ 
(GeV/c) & interactions &  Events   \\[2mm] \hline

14.5A	 &       $^{16}$O-AgBr     &  379 \\ [1mm] 
60A      &       $^{16}$O-AgBr     &  422 \\[1mm] 
200A     &       $^{16}$O-AgBr     &  223 \\[1mm] 
200A     &       $^{32}$S-AgBr     & 452 \\[1mm] \hline 
       
\end{tabular}
\caption[]{Number of events selected for the analysis.}
\end{center}
\end{table}
\end{footnotesize}

\noindent In order to compare the findings of the present work with the predictions of Monte Carlo model, \amm\cite{29,30}, matching numbers of events equal to the experimental ones are simulated using the codes \amm-v1.21-v2.21. This is done by taking into account the percentage of interactions occurring in collisions of projectile with various targets in emulsion\cite{2,26} while generating the \am events, the values of impact parameter are so set that the mean multiplicities of relativistic charged particles, $<N_{s}>$ match with those estimated for the corresponding real data samples. Furthermore, in order to test whether the fluctuations in some of the observables characterizing an event are of dynamic origin, the findings are compared with those obtained from the analysis of data samples which are free from the dynamical fluctuations. The technique of event mixing gives such a reference data in which dynamical correlations amongst the particles are completely destroyed. The mixed event samples corresponding to the real and \am samples are generated by adopting the standard procedure of event mixing, as described in \cite{2,27,31}.\\

\section{Analysis and Results}

\noindent Mean values of pseudorapidity on ebe basis are determined as,
\begin{eqnarray}
              M_{\eta}=\frac{1}{N}\sum_{i=1}^{N}\eta_{i}
\end{eqnarray}
\noindent where, N is the number of charged particles recorded in each event and having their $\eta$ values in the range $\eta \pm 3.0 $. Comparison of $M_{\eta}$ distributions for various data sets, real and \amm, at different energies with their corresponding mixed events are exhibited in Fig.1. It is interesting to note in the figures that the distributions for the real data do have a long tail as compared to those due to the mixed events, indicating the presence of fluctuations other than the statistical ones. The tails, however, are noticed to be more pronounced at lower energies. Distributions representing the \am events too exhibit similar trends but with somewhat smaller magnitudes as compared to the experimental data. Mean values of \(M_\eta \) and dispersion, \(\sigma_\eta\) of \(M_\eta \)-distributions for various data sets are estimated and presented in Table~2. Values within brackets are due to the corresponding mixed events. It may be noted from Fig.1 and Table~2 that the distributions become narrower and shift towards higher values of \(M_\eta\) and the values of \(< M_\eta >\) are noticed to become higher with increasing beam energy and system size\cite{12}.\\

\noindent In order to quantify the amount of deviation of fluctuations from the one expected on the basis of statistically independent particle emission, the magnitude of the fluctuations, $\omega_{\eta}$ in the quantity $M_{\eta}$ is defined as,
\begin{eqnarray}
              \omega_{\eta}=\frac{<M_{\eta}^{2}> - <M_{\eta}>^{2}}{<M_{\eta}>} = \frac{\sigma_{M_{\eta}}}{<M_{\eta}>}
\end{eqnarray}
\noindent and the difference, $d$ in the values of $\omega_{\eta}$ for the data and mixed events distributions,
\begin{eqnarray}
              d = \omega_{\eta}(data) - \omega_{\eta}(mixed)
\end{eqnarray}
\noindent gives the difference in fluctuations from the random baseline. A positive value of $d$, if obtained for a given data set would indicate the presence of correlations, like Bose-Einstein correlation\cite{7,32}. Values of d for data and \am samples at different energies are presented in Table$~$2. It may be noted that the values of $d$ are positive for all the data sets.\\

\noindent For checking the compatibility between the data and MC model, \amm, values of mean charged particle multiplicity $<N_{bin}>$, in a $\eta$ window of fixed width, $\Delta\eta$ are calculated for the real and \am data samples. Dependence of \(< N_{bin} >\) on the \(\eta\)-window width are shown in Fig.2. It may be noted in the figure that the experimental and model predicted values of \(< N_{bin} >\) are quite close irrespective of the fact that how small or large is the $\eta$ window. This, in turn encourages one to estimate the various measures of fluctuations in limited $\eta$ windows, $\Delta\eta$ and to compare with the model predictions.\\

\noindent It has been suggested\cite{33} that the comparison of real and mixed event $M_{\eta}$ distributions may not fully account for the rare non-statistical fluctuations of large amplitude and therefore, in order to quantify and study the deviations of $M_{\eta}$ distributions from the baseline, several approaches have been suggested\cite{21,22,33,34,35}. In the present analysis the method proposed by Gazdzicki and Mrowczynski\cite{21} is adopted, according to which the ebe fluctuations of observables defined as a sum of particle's kinematical variables on ebe basis such as, $\eta$, $p_{T}$, etc, is expected to lead to some interesting conclusions. It has been shown\cite{22} by studying the second moment of the distributions of such kinematical variables that it may be possible to evaluate the degree of randomization and thermalization characteristics of high multiplicity events produced in AA collisions.\\

\noindent As described in refs.21 and 22, for every particle in a given event, a quantity,
\begin{eqnarray}
              z_{i}=\eta_{i} - <\eta>
\end{eqnarray}
\noindent is defined, where $<\eta>$ denotes the mean value of $\eta$ distributions of the entire sample of event. Then for the variable $z_{i}$, of each event, the variable Z is calculated as,
\begin{eqnarray}
              Z = \Sigma_{i=1}^{N_{ev}}z_{i}
\end{eqnarray}
\noindent where N$_{ev}$ is the total number of events in the data sample. \\

\noindent From these definitions, the measure $\Phi_{\eta}$ is estimated as,
\begin{eqnarray}
              \Phi_{\eta} = \sqrt{\frac{<Z^{2}>}{<N>}}-\sqrt{<z^{2}>}
\end{eqnarray}
\noindent where $<Z^{2}>$ is the second moment of the inclusive z distribution. $\Phi_{\eta}$, would, thus, quantify the degree of fluctuations in mean pseudorapidity from event to event. For independent emission of particles from a single source the value of $\Phi$ will be zero, whereas, if AA collisions are regarded as the incoherent superposition of multiple independent \(nn\) collisions, the value of $\Phi$ would match with that measured for \(nn\) collisions\cite{22}.\\

\noindent Values of \(\Phi_\eta\) for various data sets, real, \am and mixed, are estimated and their dependence on the beam energy are studied by plotting \(\Phi_\eta\) against \(lnE\) as shown in Fig.3. The values of \(\Phi_\eta\) for \(pp\) collisions at 200 GeV are also displayed in the same figure; a data sample comprising of 397 events produced in \(pp\) collisions at 200 GeV, available in the laboratory are utilized for the purpose, whereas, for \am predictions, a sample of 10$^5$ events are simulated. It is noticed in Fig.3 that \(\Phi_\eta\) increases almost linearly for AA collisions, while for \(pp\) collisions, data point falls far below than the one expected from the observed linear behaviour of \(\Phi_{\eta}\) against \(lnE\). Findings from the \am event analysis also match with this observation except that the model predicted values of \(\Phi_{\eta}\) are smaller as compared to those obtained from the real data. In order to test whether \(\Phi_{\eta}\) values depend on the number of \(nn\) collisions, the variations of \(\Phi_{\eta}\) with mean number of participating target nucleons, $<N_{part}>_{tgt}$ for the \am data are plotted in Fig.4. The value of \(\Phi_{\eta}\) for pp collisions at 200 GeV, is also  shown in the same figure. It is interesting to note that the values of $\Phi_{\eta}$, although slightly decrease with the increase of $<N_{part}>_{tgt}$, are significantly larger than those obtained for pp data samples. In case of real data, since the values of the $<N_{part}>$ are not measured, the samples are divided into subgroups according to their $N_{g}$ values, where $N_{g}$ represents the number of tracks with relative velocities lying in the range $0.3\leqslant\beta\leqslant0.7$. These tracks correspond to the number of knock-out target protons and are related to the number of \(nn\) collisions. Variations of $\Phi_{\eta}$ with $<N_{g}>$  at different energies are exhibited in Fig.5. Values of $\Phi_{\eta}$ for 200 GeV \(pp\) interactions are also displayed in the same figure.  It is evidently clear from the figure that the value of $\Phi_\eta$ for various $N_{g}$ groups are much larger than the corresponding value obtained for \(pp\) collisions. Furthermore,  the observed larger values of $\Phi_\eta$ from the real data  than those predicted by \am  generator suggest that the fluctuations of larger magnitudes are present in the real data. It should be mentioned here that the values of $\Phi_{\eta}$ observed in the case of transverse momenta and transverse energy analyses carried out by CERES and PHENIX collaborations\cite{7}, are much larger than those observed in the present analysis. This may indicate that the $\eta$-distributions are not as sensitive as the $p_{T}$ and $E_{T}$ distributions. However, since  $\Phi_{\eta}$, considered in the present study is sensitive to the long-range correlations and the values of \(\Phi_\eta\) spread throughout the full phase space, a value, \(\Phi_{\eta} > 0\) would suggest  searching for the presence of some long-range correlations or large scale clustering in the particles $\eta$-distribution.\\

\noindent Dependence of $\Phi_{\eta}$ on pseudorapidity-bin widths has also been looked into by estimating the \(\Phi_\eta\) values  in \(\eta\)-windows of limited widths. For the purpose, a \(\eta\)-window of fixed width, \(\Delta\eta\) is selected and placed in such a way that its centre coincides with the centre of symmetry of \(\eta\) distribution, \(\eta_c\). Starting from a window of width, \(\Delta\eta\) = 0.5, its width is increased in steps of 0.5 untill the entire \(\eta\) region considered, i.e. \(\Delta\eta\) = \(\pm\) 3.0, is exhausted. Shown in Fig.6 are the variations of \(\Phi_\eta\) with \(\Delta\eta\) for various event samples at different energies considered. It is interesting to note in the figure that for AA collisions,  the values of \(\Phi_\eta\) increase with \(\Delta\eta\) in a regular fashion and for a given \(\eta\)-window \(\Phi_\eta\) increases with incident beam energy and projectile mass. The values of this parameter for the mixed events are nearly zero irrespective of \(\eta\)-window width or type of collisions. It is also interesting to notice in the figure that the values of \(\Phi_{\eta}\) for pp collisions at 200 GeV are much smaller as compared to those obtained for $^{16}$O-AgBr or  $^{32}$S-AgBr collisions.\\

\noindent The presence of jet-like phenomenon or cluster production and determination of their sizes and frequency has been investigated by following the algorithm applied to $p\bar{p}$ collisions\cite{2,36}, which is somewhat different from the approach adopted in refs. 37 and 38, where cluster production and their sizes were searched for by histogramming the pseudorapidity differences amongst the $n^{th}$ nearest neighbours. The present algorithm is rather more suitable for identifying the high density regions in $\eta-\phi$ space, which provides a clean separation in the $\eta-\phi$ metric in the low multiplicity and low particle density in the final state\cite{2,22}. The investigations involving jet-like phenomenon are suitable for high particle density data as well as for understanding the degree of clustering in two dimensional $\eta-\phi$ space. This method is envisaged to help estimate the cluster frequencies, cluster multiplicities and fractions of particles produced through cluster decays on ebe basis. The cluster structure of multiparticle final states produced in AA collisions involves the concept of Jaynes Shannon entropy. As the observables are very sensitive to the total event multiplicities, comparison of results from real and mixed event analysis might lead to some useful information. A detailed description of the analysis technique which involves grouping of particles into clusters on ebe basis has been presented elsewhere\cite{12,39,40}, however considering worthwhile, a brief description of analysis procedure is presented below.\\

\noindent For a particle i of an event with multiplicity k in the considered $\eta-\phi$ space, its $r_{ik}$ values with respect to the next particle k (k$\neq$i) is estimated using the relation $r_{ik} = \sqrt{(\delta\phi^{2} + \delta\eta^{2})}$, where $\delta\phi$ and $\delta\eta$ respectively represent the differences in azimuthal angles and pseudorapidities of $i^{th}$ and $k^{th}$ particles. This gives a cone of radius $r_{ik}$ which contains i and k particles. Thus, starting from $i = 1$, i.e. from first particle, its $r_{ik}$ values are calculated with respect to $(i+k)^{th}$ particle and if this value is less than a pre-fixed value r, the particle is added to form a cluster. A cluster is taken to be genuine if it has at least m particles with $m\geq2$. Once a cluster in an event is obtained, another cluster is searched for using the remaining particles of the event. After the cluster identification, the following parameters are estimated for a given value of pre-fixed cone of  radius r :
\begin{enumerate} 
\item[i.]  fraction of particles produced through cluster, with each cluster having at least m particles
\item[ii.] number of clusters per event, or the cluster frequency and
\item[iii.]entropy, $S = \Sigma_{k}p_{k}\ln{p_{k}}$, where the summation was over all clusters with multiplicities $m\geq2$. The quantity $p_{k} = n_{k}/\Sigma_{k}n_{k}$ is the probability of finding a particle in the $k^{th}$ cluster\cite{40}.
\end{enumerate}
\noindent The Jaynes Shannon entropy, introduced in this way, is regarded as a good measure of the ``amount of uncertainty'' represented by a discrete probability distribution. It is, therefore, expected that it would help distinguish between different heavy-ion collisions\cite{40}. It should be noted that for a very small value of r, there may be no or only a few clusters in an event, while for significantly larger values of r, almost all the particles belonging to an event will be grouped in a single cluster\cite{2}.\\

\noindent Considering a cluster with multiplicity, m $\geq 5$, number of clusters on ebe basis, fraction of particles produced through clusters and entropy, S are calculated for each data sample. Variations of mean number of clusters with \(r^2\) for the real and MC data sets are plotted in Fig.7, whereas dependence of S on \(r^2\) is shown in Fig.8. It may be noted from these figures that there are clear peaks at the lower values of \(r^2\) and thereafter, values of \(<N_{cl}>\) or S decreases quickly first and then tend to acquire nearly a constant value beyond \(r^2 \sim 2\). \am data too show similar trends of variations of \(<N_{cl}>\) and  S with \(r^2\) but with somewhat smalller magnitudes. However, for the mixed events the maximum occurs at relatively higher values of \(r^2\) and the patterns are rather quiet. It may also be noted in these figures that the maxima shift towards the lower values of \(r^2\) with increasing beam energy or system size. In order to have a clear reflection of the dominance of clusterization in the data as compared to correlation free MC events, ratio of maximum value of \(<N_{cl}>\) and S with these in the last bin of \(r^2\), \(\rho_c\) and \(\rho_s\) are calculated for cluster size, m = 5 and presented in Table~3 and Table~4. It may be noted from the table that the values of \(\rho_c\) and \(\rho_s\) increase with the beam energy and size of the colliding nuclei and are much larger than those obtained from the corresponding mixed event sets.\\

\noindent Variations of fraction of particles produced through the cluster decays with \(r^2\) for \(m \geq 5\) are displayed in Fig.9. It may be noticed in these figures that for \(r^2 < 1\), nearly 55-70\% of particles are produced through the formation of clusters with multiplicity \(\geq\) 5. These numbers are found to be much larger than those estimated from the analysis of the corresponding mixed event samples. The \am model too predicts the presence of nearly similar clustering effects. These observations, thus, tend to suggest that clusters of larger sizes are formed as the energy or the size of the colliding nuclei increases. A comparison of these findings with those due to the mixed events indicates the presence of dominant clustering effect in the real data which is nicely supported by the predictions of \am model. These findings thus, do not support the hypothesis of independent particle emission.

\section{Summary}

Event-by-Event fluctuations in mean pseudorapidity of charged particles produced in the full rapidity space are examined and compared with the base line distributions. It is observed that the $M_{\eta}$ distributions exhibit a relatively larger tails as compared to the corresponding baseline distributions. Moreover, the distributions become narrower and shift towards the higher values of $\eta$ with increasing energy or system size. The fluctuation strength measure, $\Phi_{\eta}$ is found to increase almost linearly, if plotted against $lnE$ for AA data which is nicely supported by \am model except that the model predicted values are relatively smaller. However the deviation of pp data from the linearity indicates that AA collisions can not be described by simple superposition of multiple nn collisions. This observation is further supported from the observations of $\Phi_{\eta}$ dependence on the number of participating nucleons, where $\Phi_{\eta}$ is observed to decrease with $<N_{part}>_{tgt}$ or $<N_{g}>$. Presence of cluster-like multihadron final states were searched for in two dimensional $\eta-\phi$ space by applying the cone algorithm. The findings indicate that clusters of different sizes are present in the data and with increasing beam energy or system size more and more particles tend to be grouped in a cluster.


\newpage
\begin{footnotesize}
\begin{table}
\begin{center}
\begin{tabular}{|c|c|c|c|c|c|c|} \hline
&	Type of        &		Energy	   &   &    &    &   \\
&	Interaction    &    (A GeV/c)  &   $<M_{\eta}>$         &  $\sigma$(M$_{\eta}$)  &   $\omega_{\eta}$    	       &   d \\ [2ex]  \hline
&	$^{16}$O-AgBr  &    14.5       &  2.09 $\pm$ 0.02       &  0.48 $\pm$ 0.02       &  0.231 $\pm$ 0.008        &  0.062\\
	&        &  		  &  \footnotesize{(1.97 $\pm$ 0.02)} &  \footnotesize{(0.33 $\pm$ 0.01)} & \footnotesize{(0.169 $\pm$ 0.006)}  &       \\ [1mm]
&   $^{16}$O-AgBr  &  	60         &  2.61 $\pm$ 0.02       &  0.34 $\pm$ 0.01       &  0.131 $\pm$ 0.004        &  0.047\\
	Expt.     &        &    &  \footnotesize{(2.52 $\pm$ 0.01)} & \footnotesize{(0.21 $\pm$ 0.01)}  & \footnotesize{(0.084 $\pm$ 0.003)}  &   \\ [1mm]
&   $^{16}$O-AgBr  & 	200  &  3.08 $\pm$ 0.02       &  0.24 $\pm$ 0.01       &  0.079 $\pm$ 0.004       &  0.003\\
	&       &       	 & \footnotesize{(3.13 $\pm$ 0.01)}  & \footnotesize{(0.24 $\pm$ 0.01)}  & \footnotesize{(0.076 $\pm$ 0.003)}  &       \\ [1mm]
&	$^{32}$S-AgBr &  200  &  3.16 $\pm$ 0.01  &  0.31 $\pm$ 0.01  &  0.099 $\pm$ 0.003  &  0.024\\ 
	&   &     & \footnotesize{(3.13 $\pm$ 0.01)}  & \footnotesize{(0.24 $\pm$ 0.01)}  & \footnotesize{(0.075 $\pm$ 0.002)}  &       \\  [1mm] \hline 
&	$^{16}$O-AgBr   &  14.5  &  1.80 $\pm$ 0.01  &  0.28 $\pm$ 0.01  &  0.156 $\pm$ 0.006  &  0.024\\ 
	&         &  	 &  \footnotesize{(1.75 $\pm$ 0.01)}  &  \footnotesize{(0.23 $\pm$ 0.01)}  &  \footnotesize{(0.132 $\pm$ 0.005)}  &       \\ [1mm] 
&   $^{16}$O-AgBr   &  60  &  2.47 $\pm$ 0.02  &  0.24 $\pm$ 0.01  &  0.139 $\pm$ 0.005  &  0.007\\
	AMPT     &   &  &  \footnotesize{(2.37 $\pm$ 0.01)}  & \footnotesize{(0.26 $\pm$ 0.01)}  & \footnotesize{(0.109 $\pm$ 0.004)}  &       \\ [1mm]
&   $^{16}$O-AgBr   &   200  &  3.00 $\pm$ 0.01  &  0.22 $\pm$ 0.01  &  0.072 $\pm$ 0.003  &  0.025\\
	&    &  	 & \footnotesize{(2.96 $\pm$ 0.01)}  & \footnotesize{(0.14 $\pm$ 0.01)}  & \footnotesize{(0.047 $\pm$ 0.002)}  &       \\ [1mm]
&	$^{32}$S-AgBr   &   200  &  3.12 $\pm$ 0.01  &  0.11 $\pm$ 0.01  &  0.034 $\pm$ 0.001  &  0.006\\ 
	&     &   & \footnotesize{(2.77 $\pm$ 0.01)}  & \footnotesize{(0.11 $\pm$ 0.01)}  & \footnotesize{(0.028 $\pm$ 0.001)}  &       \\  [1mm]  \hline

\end{tabular}
	\caption[]{Values of $<M_{\eta}>$, \(\sigma M_{\eta}\), \(\omega_{\eta}\) and d for the experimental and \am events. Values within brakets are due to the corresponding mixed events. }
\end{center}
\end{table}
\end{footnotesize} 

\newpage
\begin{footnotesize}
\begin{table}
\begin{center}
\begin{tabular}{|c|c|c|c|} \hline
Type of       &   Energy	   & \multicolumn{2}{|c|}{$\rho_{c}$}  			\\ \cline{2-4}
Interaction   &  (A GeV/c) & \multicolumn{2}{|c|}{m = 5}       			\\ \cline{2-4}
              &            &  Expt.    			  &    Mixed             \\  \hline
$^{16}$O-AgBr &  14.5      &  1.178 $\pm$ 0.050	  & 	1.032 $\pm$ 0.048 	\\  
$^{16}$O-AgBr &  60        &	 1.431 $\pm$ 0.050	  & 	1.143 $\pm$ 0.043 	\\
$^{16}$O-AgBr &  200       &  1.984 $\pm$ 0.082	  & 	1.406 $\pm$ 0.063 	\\
$^{32}$S-AgBr &  200       &  2.661 $\pm$ 0.097	  &	1.735 $\pm$ 0.072 	\\  [1mm] \hline 
			  &            &  AMPT     			  &     Mixed        	 \\  \hline
$^{16}$O-AgBr & 	14.5       &  1.039 $\pm$ 0.041	  &   1.035 $\pm$ 0.034 	\\
$^{16}$O-AgBr & 	60         &  1.303 $\pm$ 0.058	  &   1.219 $\pm$ 0.054 	\\
$^{16}$O-AgBr & 	200        &  1.207 $\pm$ 0.052	  &   1.295 $\pm$ 0.057 	\\
$^{32}$S-AgBr & 	200        &  2.374 $\pm$ 0.068	  &   1.613 $\pm$ 0.083 	\\ [1mm]  \hline
\end{tabular}
	\caption[]{Values of \(\rho_{c}\) for the experimental and \am data samples at different energies.}
\end{center}
\end{table}
\end{footnotesize} 

\begin{footnotesize}
\begin{table}
\begin{center}
\begin{tabular}{|c|c|c|c|} \hline
Type of        &		Energy	  &	 \multicolumn{2}{|c|}{$\rho_{s}$}       \\ \cline{2-4}
Interaction    &  (A GeV/c)    &    \multicolumn{2}{|c|}{m = 5}          \\  \cline{2-4}
  			   &               &   Expt.               &   Mixed          \\  \hline
$^{16}$O-AgBr  &  	 14.5      &  1.072 $\pm$ 0.610    & 1.011 $\pm$ 0.124 \\  
$^{16}$O-AgBr  &    60         &	 1.167 $\pm$ 0.139     & 1.045 $\pm$ 0.146 \\
$^{16}$O-AgBr  &    200        &  1.306 $\pm$ 	0.910  & 1.171 $\pm$ 0.256 \\
$^{32}$S-AgBr  &    200        &  1.375 $\pm$ 	0.369  & 2.292 $\pm$ 0.687 \\  [1mm] \hline 
			   &               &       AMPT             &      Mixed          \\  \hline
$^{16}$O-AgBr  & 	14.5       &  1.018 $\pm$ 0.105     &   1.002 $\pm$ 0.098 \\
 $^{16}$O-AgBr & 	60         &  1.129 $\pm$ 0.204     &   1.100 $\pm$ 0.187 \\
 $^{16}$O-AgBr & 	200        &  1.055 $\pm$ 0.212     &   1.098 $\pm$ 0.216 \\
$^{32}$S-AgBr  & 	200        &  1.394 $\pm$ 0.202     &   1.236 $\pm$ 0.347 \\ [1mm]  \hline
\end{tabular}
\caption[]{Values of \(\rho_{s}\) for the various data sets, real and \am at different energies.}
\end{center}
\end{table}
\end{footnotesize} 

\newpage
\begin{figure}[th]
  \includegraphics[width=\linewidth]{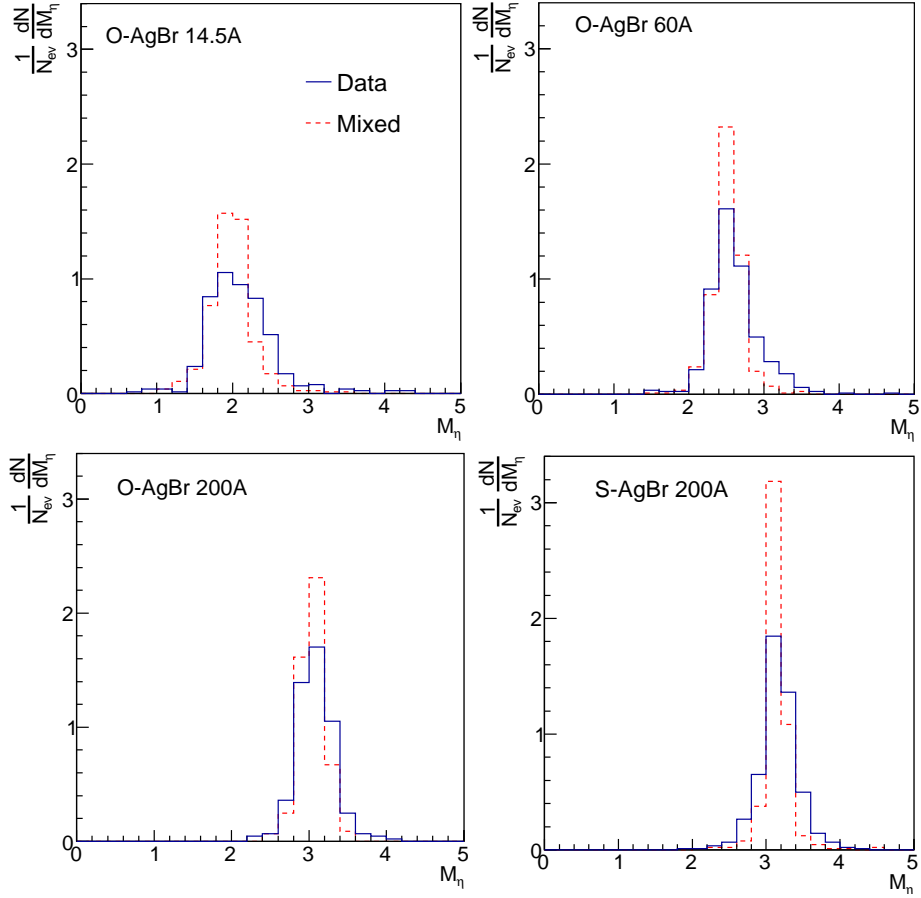}
  \caption{ Comparison between the data and mixed event $M_{\eta}$-distributions for the experimental and \am events.}
\end{figure}
\newpage
\begin{figure}[th]
  \includegraphics[width=\linewidth]{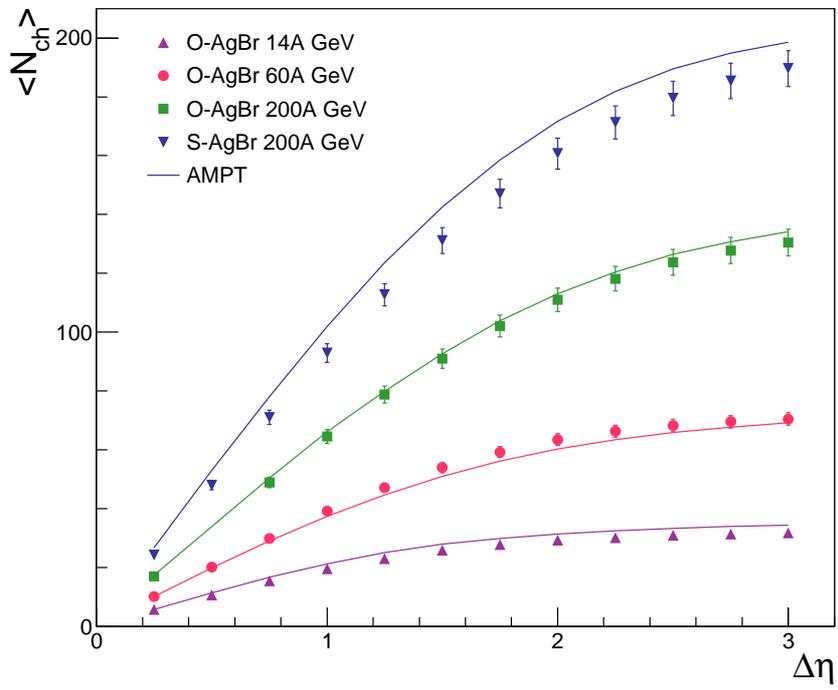}
  \caption{ Dependence of mean charged particle multiplicities on \(M_\eta\)-bin widths for the real and \am events.}
\end{figure}
\newpage
\begin{figure}[th]
  \includegraphics[width=\linewidth]{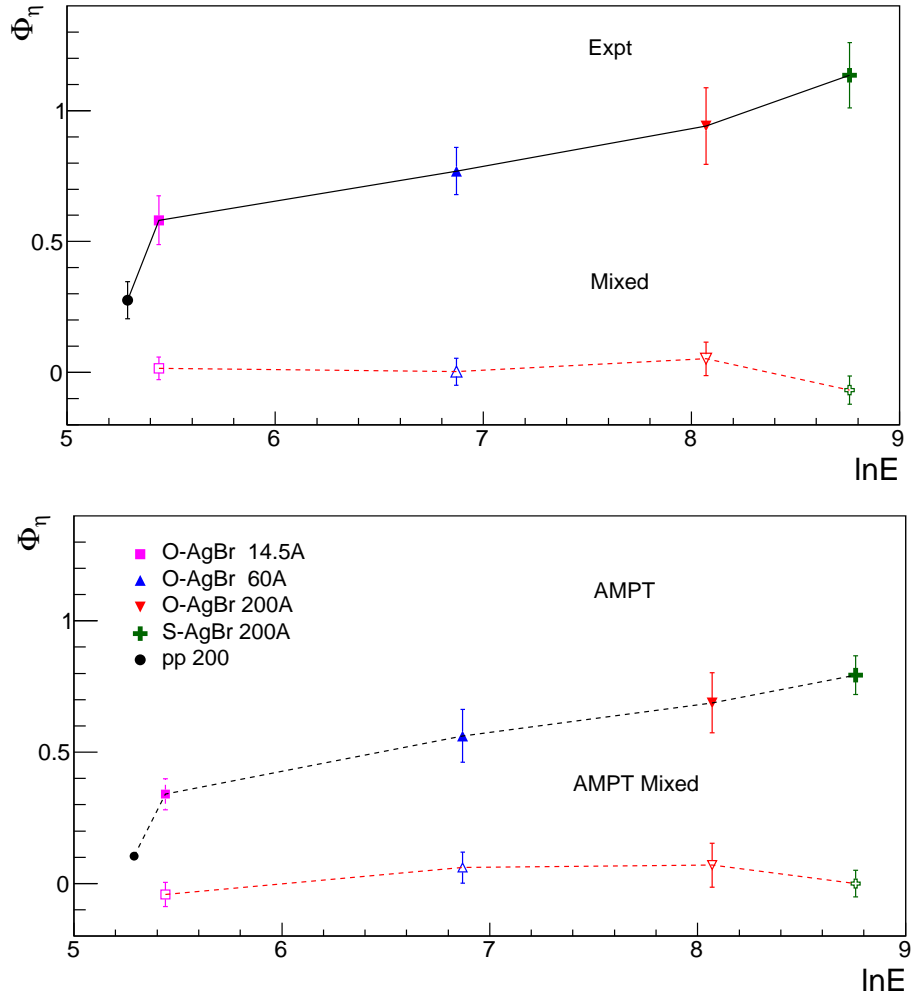}
	\caption{\(\Phi_\eta\) dependence on \(lnE\). Symbols with open markers are due to the mixed events corresponding to the respective Expt and \am event samples.}
\end{figure}
\newpage
\begin{figure}[th]
  \includegraphics[width=\linewidth]{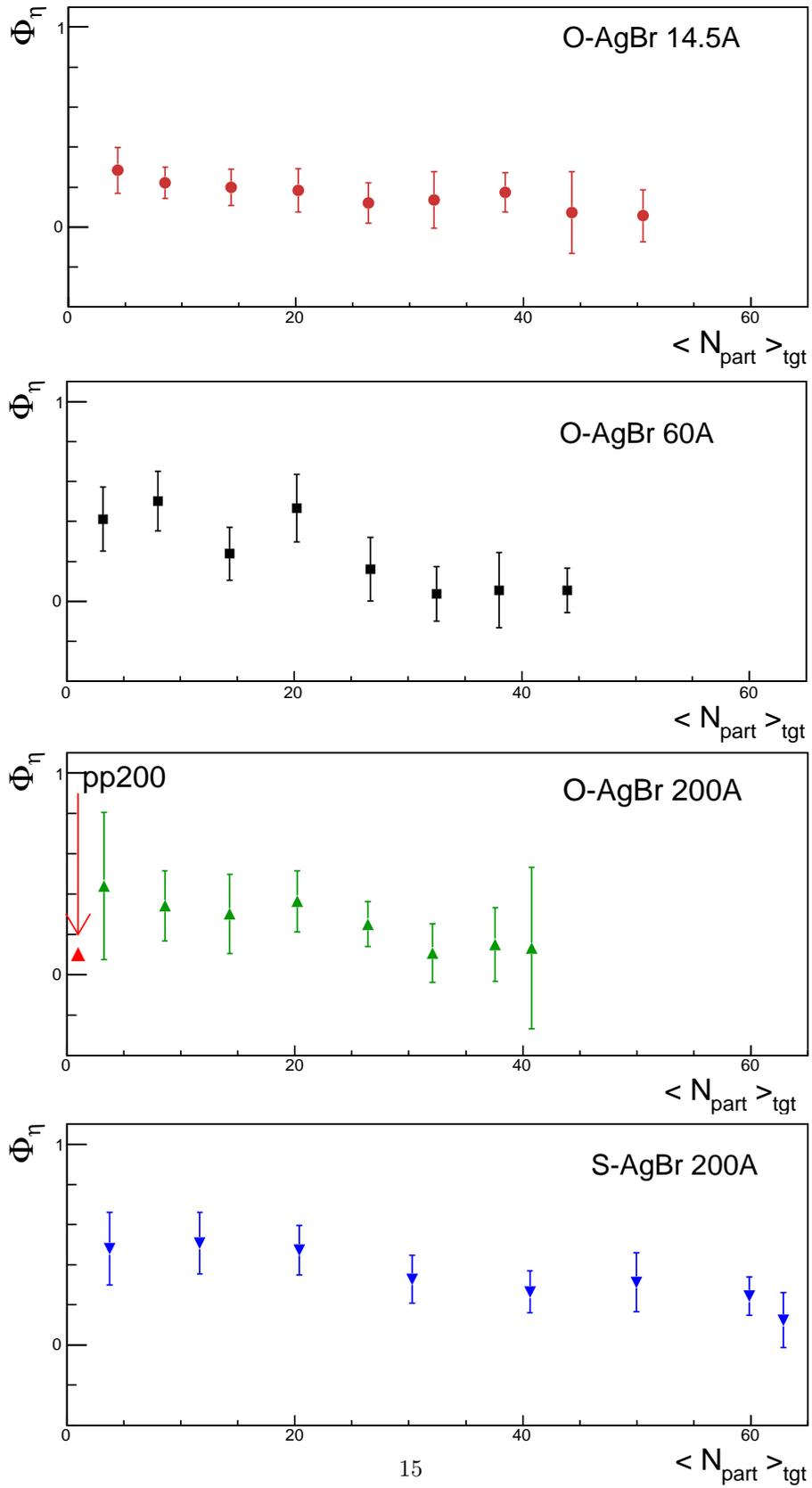}
\caption{Variations of \(\Phi_\eta\) with mean number of target participants, \(< N_{part} >_{tgt}\) for the \am events.}
\end{figure}
\newpage
\begin{figure}[th]
  \includegraphics[width=\linewidth]{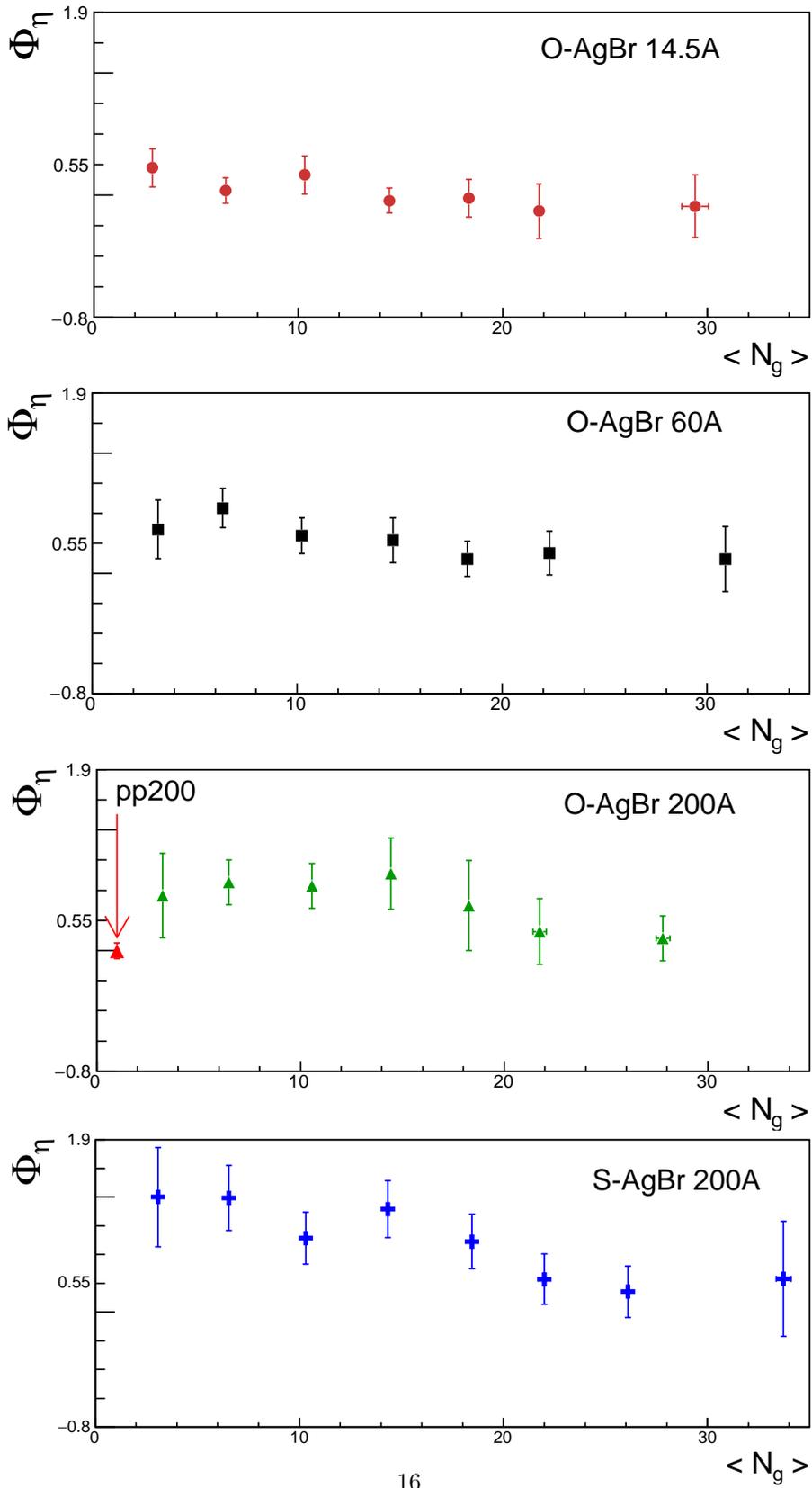}
	\caption{Variations of \(\Phi_\eta\) with \(< N_g >\) for the experimental data.}
\end{figure}
\newpage
\begin{figure}[th]
  \includegraphics[width=\linewidth]{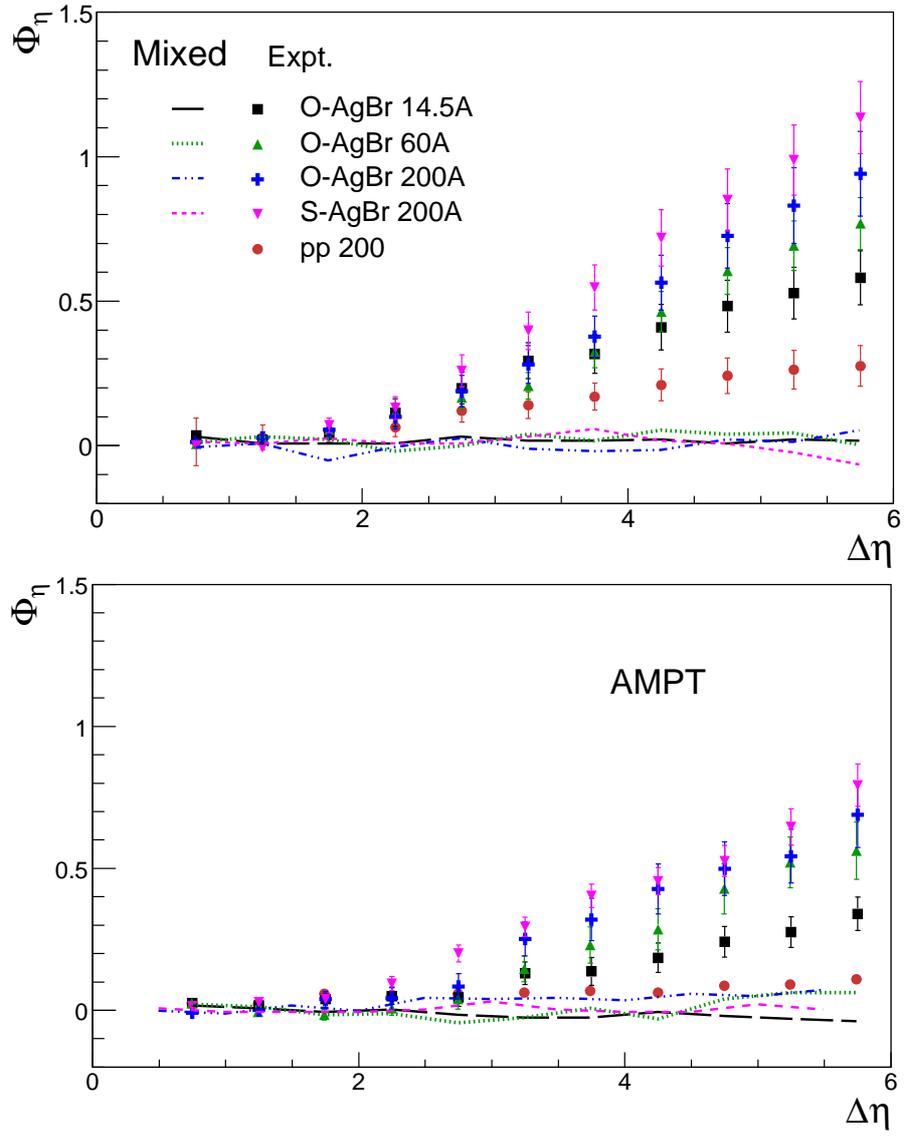}
\caption{Dependence of \(\Phi_\eta\) on \(\Delta\eta\) for the real and MC events. The lines are due to the mixed event sample.}
\end{figure}
\newpage
\begin{figure}[th]
  \includegraphics[width=\linewidth]{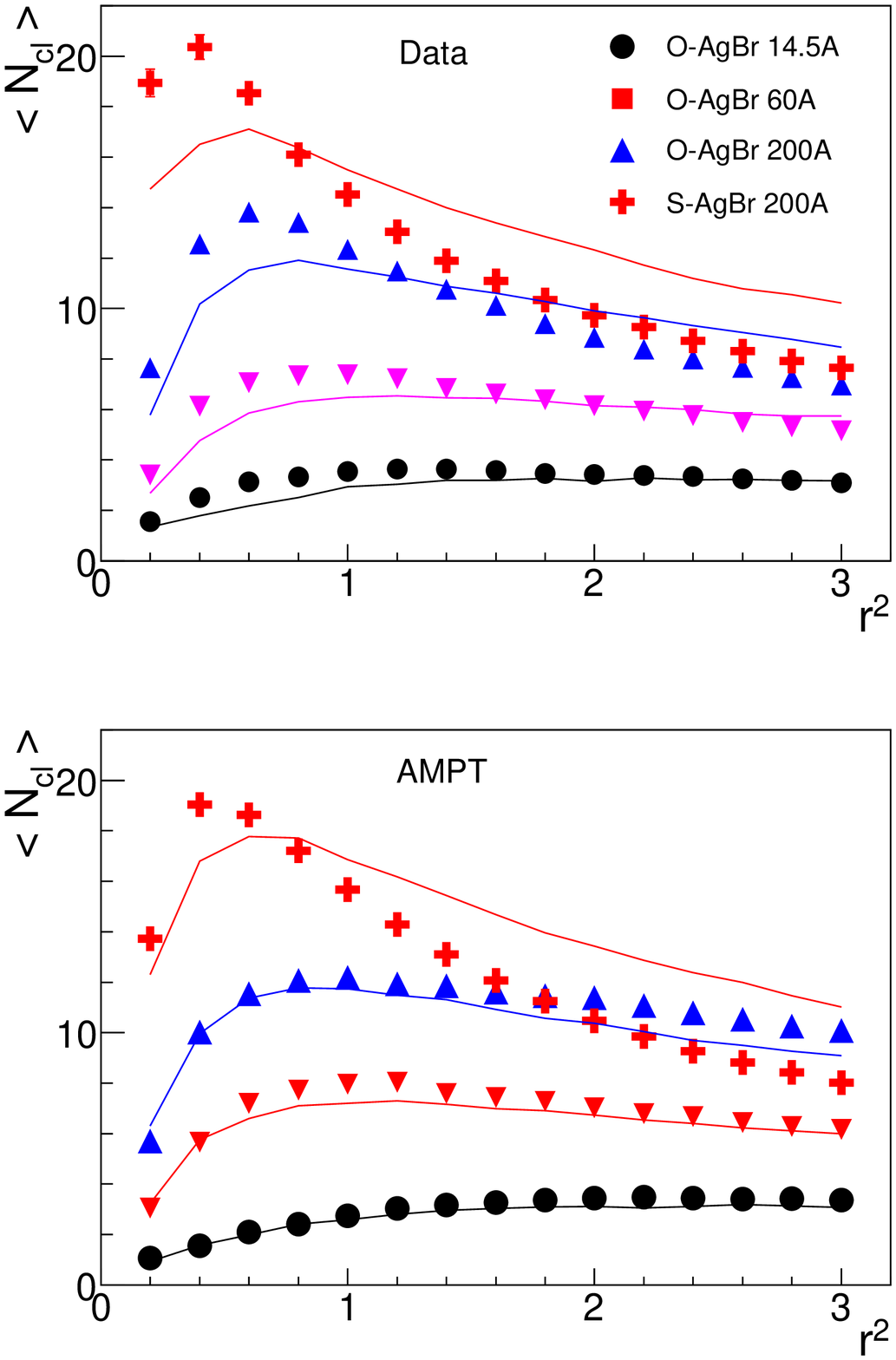}
\caption{Variations of \(<N_{cl}>\) with \(r^2\) for the real, \am and mixed events. Points correspond to real or \am events, while the lines are due to the corresponding mixed events.}
\end{figure}
\newpage
\begin{figure}[th]
  \includegraphics[width=\linewidth]{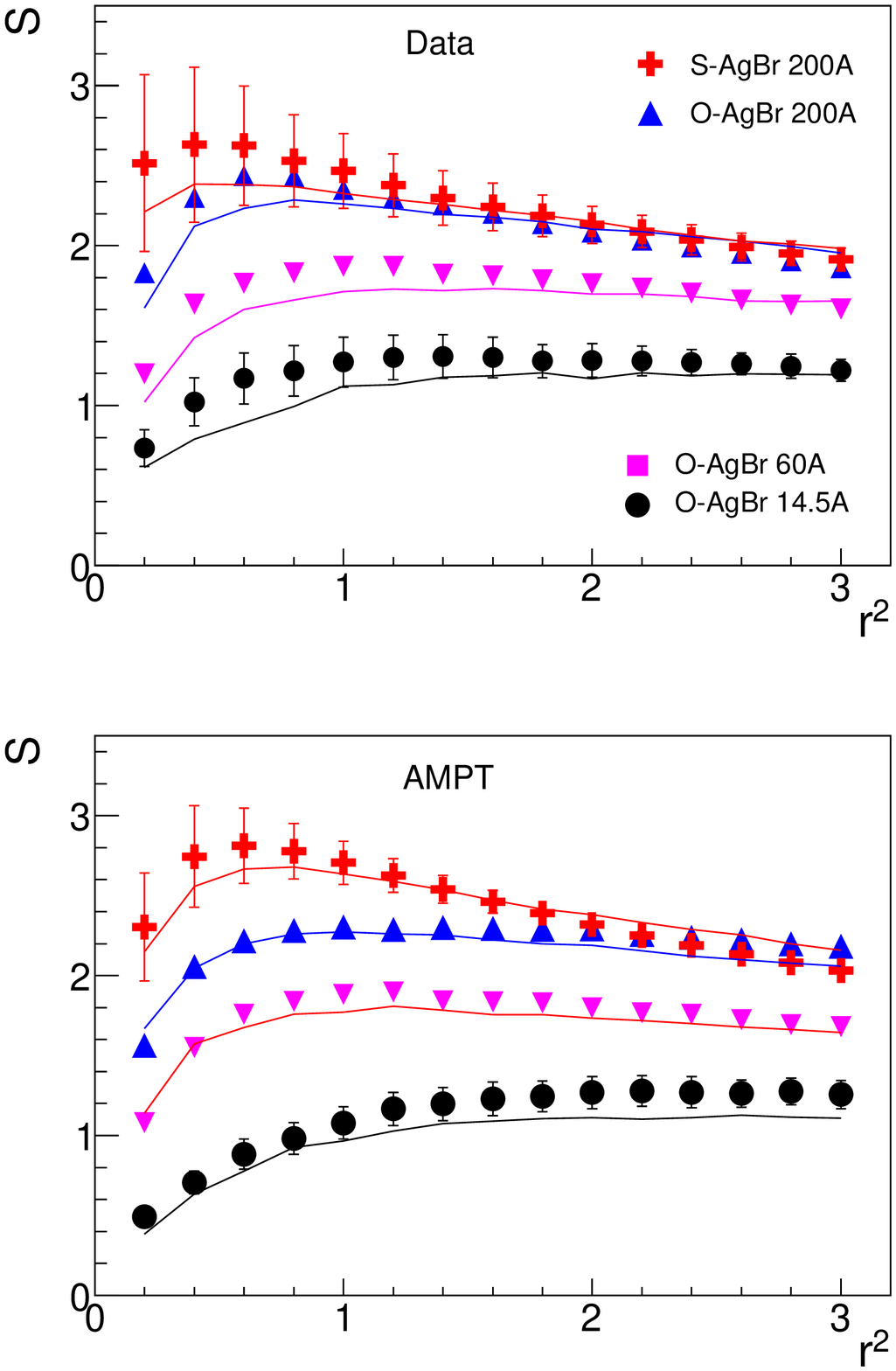}
\caption{Entropy, S vs \(r^2\) for various sets of events at different energies. Points are due to real or \am events, while the lines correspond to mixed events.}
\end{figure}

\newpage
\begin{figure}[th]
  \includegraphics[width=\linewidth]{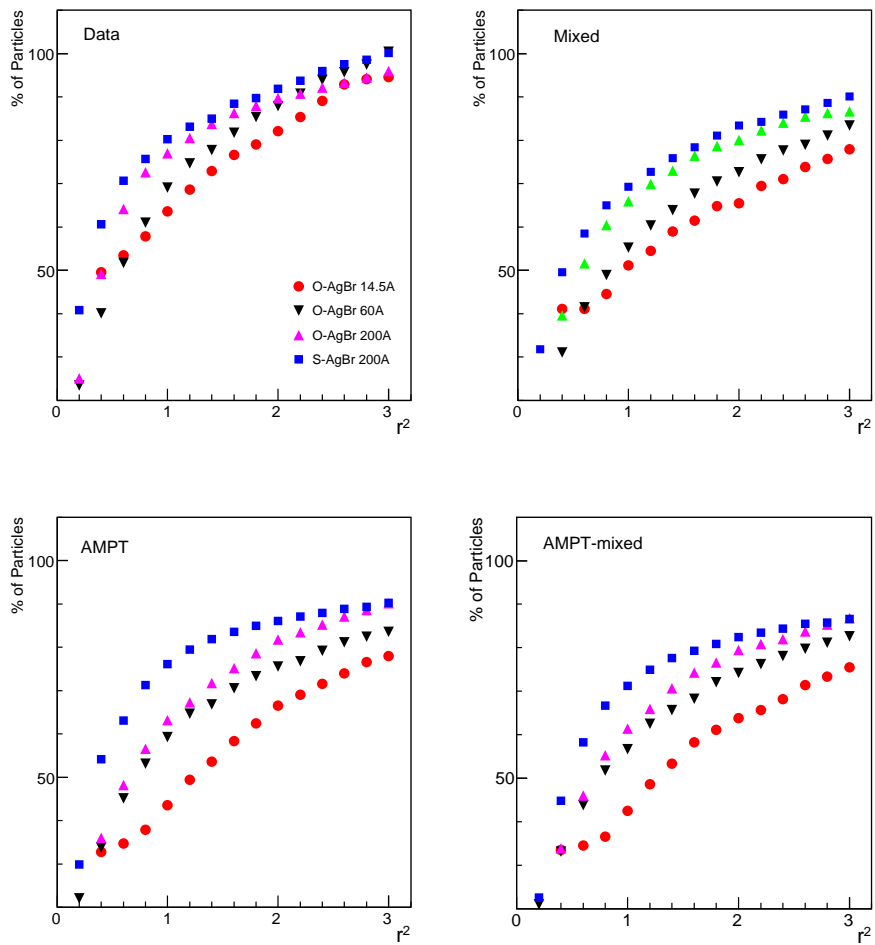}
\caption{Dependence of fraction (\%) of particles through clusters on \(r^2\) for m = 5}
\end{figure}

\end{document}